\documentclass[12pt]{article}

\usepackage[margin=2.5cm]{geometry}

\usepackage[english]{babel}
\usepackage[T1]{fontenc}
\usepackage[]{times}
\usepackage{amsmath,amssymb,amsthm,stmaryrd}
\usepackage{bbm}
\usepackage{natbib}
\usepackage{graphicx}
\usepackage{comment}
\usepackage{longtable,dcolumn,booktabs,mathtools}
\usepackage{url}

\usepackage{caption}
\usepackage{dcolumn}
\newcolumntype{d}[1]{D{.}{.}{#1}}

\newcommand{\real}{\mathbb{R}}

\newcommand{\E}{\mathbb{E}}

\newcommand{\hsp}{{\hspace{0.5mm}}} 
\newcommand{\hhsp}{{\hspace{0.1mm}}} 

\usepackage{natbib}

\begin{document}

\title{Ensemble model output statistics for wind vectors} 

\author{Nina Schuhen, Thordis L.~Thorarinsdottir and Tilmann Gneiting \\
        {\em Institute of Applied Mathematics, University of Heidelberg, Germany}} 

\date{\today}

\maketitle

\begin{abstract}
A bivariate ensemble model output statistics (EMOS) technique for the
postprocessing of ensemble forecasts of two-dimensional wind vectors
is proposed, where the postprocessed probabilistic forecast takes the
form of a bivariate normal probability density function.  The
postprocessed means and variances of the wind vector components are
linearly bias-corrected versions of the ensemble means and ensemble
variances, respectively, and the conditional correlation between the
wind components is represented by a trigonometric function of the
ensemble mean wind direction.  In a case study on 48-hour forecasts of
wind vectors over the North American Pacific Northwest with the
University of Washington Mesoscale Ensemble, the bivariate EMOS
density forecasts were calibrated and sharp, and showed considerable
improvement over the raw ensemble and reference forecasts, including
ensemble copula coupling.
\end{abstract}

\section{Introduction}

The past two decades have seen a change of paradigms in weather
forecasting, in that ensemble prediction systems have been developed
and implemented operationally \citep{Leutbecher2008}.  Ensemble
systems seek to reflect and quantify sources of uncertainty in
numerical weather forecasts, such as imperfections in initial
conditions and incomplete mathematical representations of the
atmosphere.  Despite the ubiquitous positive spread-skill relationship
\citep{Whitaker1998, Grimit2002}, ensemble forecasts tend to be
biased, and typically they are underdispersed \citep{Hamill1997}, in
that the ensemble spread is too small to be realistic.  Furthermore,
differing spatial resolutions of the forecast grid and the observation
network may need to be reconciled.

To address these shortcomings, various techniques for the statistical
postprocessing of ensemble model output have been developed
\citep{WilksHamill2007}, with ensemble model output statistics (EMOS)
or nonhomogeneous Gaussian regression \citep{Gneiting2005,
  Thorarinsdottir2010} being a state of the art method.  The EMOS
technique transforms a raw ensemble forecast into a predictive
probability density function, and simultaneously corrects for biases
and dispersion errors.  EMOS methods have been developed for
temperature and surface pressure \citep{Gneiting2005, Hagedorn2008,
  Kann2009}, where the predictive density is normal and the method is
often referred to as nonhomogeneous Gaussian regression, for
quantitative precipitation \citep{Wilks2009}, and for wind speed
\citep{Thorarinsdottir2010, Thorarinsdottir2011}.  In all these
implementations, the predictive density applies to a univariate
weather quantity.

\begin{figure}[t]
\centering
\includegraphics[width=0.75\textwidth]{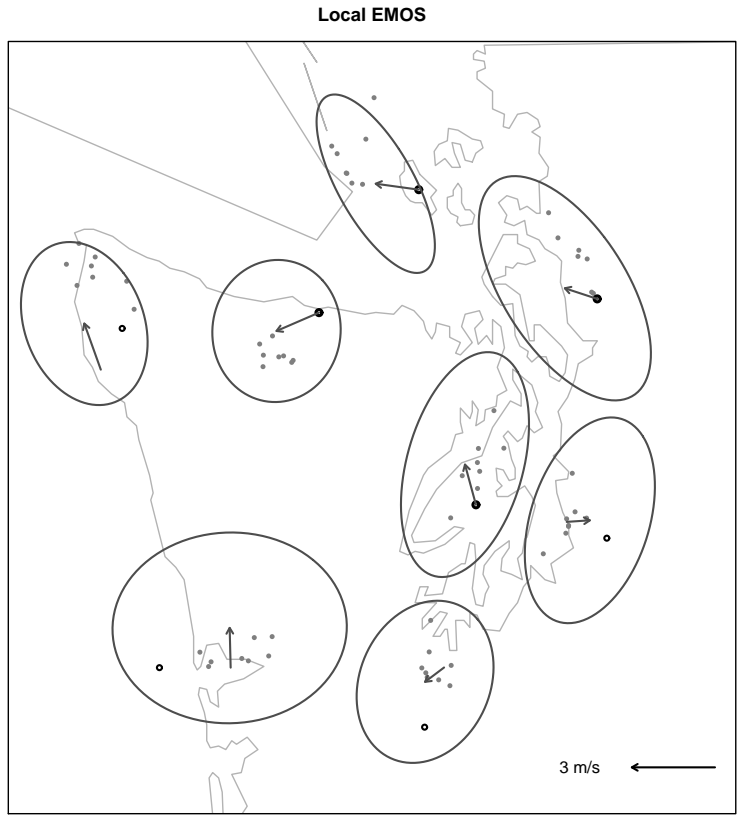}
\vspace{-7mm}
\caption{Raw ensemble and EMOS postprocessed forecasts of surface wind
  vectors at stations in the Olympic Peninsula and Puget Sound area in
  the US state of Washington, valid October 20, 2008 at 00 UTC, at a
  prediction horizon of 48 hours.  The eight members of the University
  of Washington Mesoscale Ensemble (UWME; Eckel and Mass 2005) are
  shown as gray dots.  The 75\% prediction ellipse and the mean vector
  for the EMOS density forecast are shown in dark gray, and the
  verifying wind vector is represented by a small black
  circle.  \label{fig:PugetSound}}
\end{figure}

In this current paper, we propose and develop an EMOS technique for a
bivariate weather quantity, namely surface wind vectors, comprising
both zonal and meridional components or, in an alternative but
mathematically equivalent representation, wind speed and wind
direction.  Probabilistic forecasts of wind conditions are critical in
a wide range of applications, including air traffic control, ship
routing, recreational and competitive sailing, and wind energy, where
their societal and monetary value is huge \citep{Marquis2011}.  Until
very recently, wind speed and wind direction have been addressed
independently in statistical postprocessing, without taking
dependencies into account \citep{Bao2010, Sloughter2010,
  Thorarinsdottir2010, Thorarinsdottir2011}.  However, in many of the
aforementioned applications it is important to honor the full
information about the bivariate structure of the future wind vector
that is provided by the ensemble.  Thus, our EMOS postprocessed
forecasts take the form of elliptically symmetric bivariate normal
densities, as illustrated in Figure \ref{fig:PugetSound} in an
application to the University of Washington Mesoscale Ensemble (UWME;
Eckel and Mass 2005). \nocite{Eckel2005}

The remainder of the paper is organized as follows.  In Section
\ref{sec:EMOS} we provide the details of the bivariate EMOS technique.
A case study is presented in Section \ref{sec:case.study}, where we
consider 48-hour ahead forecasts of wind vectors over the North
American Pacific Northwest in 2008 based on the eight-member UWME.
The paper closes in Section \ref{sec:discussion}, where we hint at
future developments and discuss the similarities and differences
between our EMOS technique, the BMA approach of \citet{Sloughter2009,
  Sloughter2011}, ensemble copula coupling (ECC; Schefzik 2011) and
the postprocessing method proposed by \citet{Pinson2011}, all of which
are directed at the bivariate postprocessing of ensemble forecasts of
wind vectors. \nocite{Schefzik2011} The Appendix describes our
verification methods.

\section{Ensemble model output statistics for wind vectors}  \label{sec:EMOS}

A wind vector is determined by wind speed and wind direction, or by
its zonal (west-east) and meridional (north-south) components, which
we denote by $u$ and $v$, respectively.  We now develop an ensemble
model output statistics (EMOS) method for wind vectors, where the
postprocessed probabilistic forecast takes the form of a bivariate
probability density function.  The method is tailored to ensembles
with relatively few members, such as the eight-member University of
Washington Mesoscale Ensemble (UWME; Eckel and Mass 2005), and we
illustrate it using forecast and observation data from this ensemble.

\subsection{Bivariate normal distribution}  \label{sec:BVN}

Our EMOS postprocessed forecast takes the form of an elliptically
symmetric, bivariate normal probability density function for the wind
vector $(u,v)$, with the parameters of this distribution being
specified in terms of the ensemble forecast.  The analytic form of the
bivariate normal probability density function is 
\begin{eqnarray}  
f(u,v) = 
\frac{1}{2 \pi \hhsp \sigma_u \hhsp \sigma_v \sqrt{1 - \rho_{uv}^2}} && \label{eq:BVN} \\
&& \mbox{} \hspace{-46.5mm} \times
\exp \! \left( - \frac{1}{2 \hhsp (1 - \rho_{uv}^2)}
        \left( \frac{ (u - \mu_u)^2}{\sigma_u^2}  
                - 2 \hhsp \rho_{uv} \frac{(u - \mu_u) \hhsp (v - \mu_v)}{\sigma_u \hhsp \sigma_v}  
                + \frac{(v - \mu_v)^2}{\sigma_v^2} \right) \right) \! .  \nonumber 
\end{eqnarray}
The task now is to specify the five parameters in (\ref{eq:BVN}),
namely the mean values, $\mu_u$ and $\mu_v$, of the wind vector
components $u$ and $v$, the corresponding marginal variances,
$\sigma_u^2$ and $\sigma_v^2$, respectively, and the correlation
coefficient, $\rho_{uv}$, between the wind components, in their
dependence on the ensemble forecast.

Bivariate normal density forecasts for wind vectors have also been
proposed by \citet{Gneiting2008a}, though in very crude form.  The
ingenious method of \citet{Pinson2011} estimates a dilation and
translation of an ensemble forecast of wind vectors based on bivariate
normal densities, and our approach is very similar in its treatment of
the mean and variance parameters.  However, major differences between
the approach of \citet{Pinson2011} and our method lie in the form of
the postprocessed forecast, which is a probability density function in
our case, rather than a dilated and translated ensemble, and in the
explicit modeling of the correlation coefficient $\rho_{uv}$ in our
method.  For a more detailed comparison and recommendations for a
judicious choice of the most appropriate method, given any particular
ensemble and task at hand, we refer to Section \ref{sec:discussion}.

In the description that follows, we consider a general ensemble with
$m$ members, and denote the individual wind vector forecasts by
$(u_1,v_1), \ldots, (u_m, v_m)$, respectively.

\subsection{Means}  \label{sec:mean}

In our standard implementation of the bivariate EMOS technique, the
means $\mu_u$ and $\mu_v$ are bias-corrected versions of the
respective ensemble means, in that
\begin{equation}  \label{eq:mean} 
\mu_u = a_u + b_u \bar{u}  \qquad \mbox{and} \qquad \mu_v = a_v + b_v \bar{v},
\end{equation}
where $\bar{u} = \frac{1}{m} \sum_{i=1}^m u_i$ and $\bar{v} =
\frac{1}{m} \sum_{i=1}^m v_i$.  The bias correction parameters $a_u$,
$b_u$, $a_v$ and $b_v$ are estimated from training data, in ways
described below.  In a slightly more ambitious implementation, the
mean components $\mu_u$ and $\mu_v$ are affine functions of the
individual ensemble member forecasts, namely
\[
\mu_u = a_u + b_{u,1} u_1 + \cdots + b_{u,m} u_m
\quad \mbox{and} \quad 
\mu_v = a_v + b_{v,1} v_1 + \cdots + b_{v,m} v_m, 
\]
where the regression parameters $a_u$, $b_{u,1}, \ldots, b_{u,m}$,
$a_v$ and $b_{v,1}, \ldots, b_{v,m}$ are estimated from training data.
This general version applies to ensembles with non-exchangeable
members only and reduces to the standard version when $b_{u,1} =
\cdots = b_{u,m} = \frac{1}{m}$ and $b_{v,1} = \cdots = b_{v,m} =
\frac{1}{m}$.  In our experiences with the UWME, which has
non-exchangeable members, the general version gave only very slightly
improved predictive performance, and so we report results for the
standard implementation (\ref{eq:mean}) only.

\subsection{Variances}  \label{sec:variance}

We specify the marginal variances $\sigma_u^2$ and $\sigma_v^2$ of the
bivariate normal density forecast (\ref{eq:BVN}) as affine functions
of the respective ensemble variances, in that
\begin{equation}  \label{eq:variance} 
\sigma_u^2 = c_u + d_u \hhsp s_u^2 
\qquad \mbox{and} \qquad 
\sigma_v^2 = c_v + d_v \hhsp s_v^2,  
\end{equation}
where $s_u^2 = \frac{1}{m} \sum_{i=1}^m (u_i - \bar{u})^2$ and $s_v^2
= \frac{1}{m} \sum_{i=1}^m (v_i - \bar{v})^2$.  The dispersion
correction parameters $c_u$, $d_u$, $c_v$ and $d_v$ are estimated from
training data, as described below.  To guarantee the nonnegativity of
the variances, we constrain the parameters to be nonnegative, using
the technique described by \citet{Thorarinsdottir2010}.

\subsection{Correlation coefficient}  \label{sec:corr}

The key characteristic and major innovation in our work is the
explicit modeling of the correlation coefficient $\rho_{uv}$ of the
postprocessed bivariate normal density forecast (\ref{eq:BVN}).

\begin{figure}[t]
\centering
\includegraphics[scale=0.275]{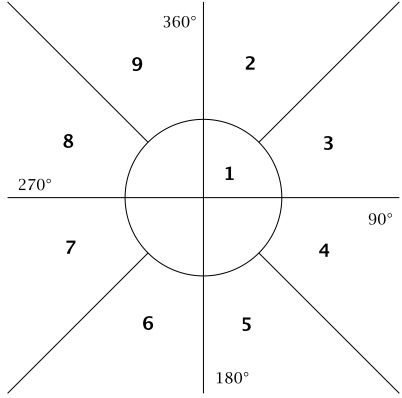}
\caption{Sectors for wind vector forecasts and observations in the
  $(u,v)$ plane.  Sector 1 is the circular region that is centered at
  the origin and corresponds to wind speeds less than or equal to two
  meters per second.  Sectors 2--9 are assigned clockwise, with
  Sectors 2--3 corresponding to a south-westerly, Sectors 4--5 a
  north-westerly, Sectors 6--7 a north-easterly, and Sectors 8--9 a
  south-easterly wind.  \label{fig:sections}}
\end{figure}

\begin{figure}
\centering
\includegraphics[width=\textwidth]{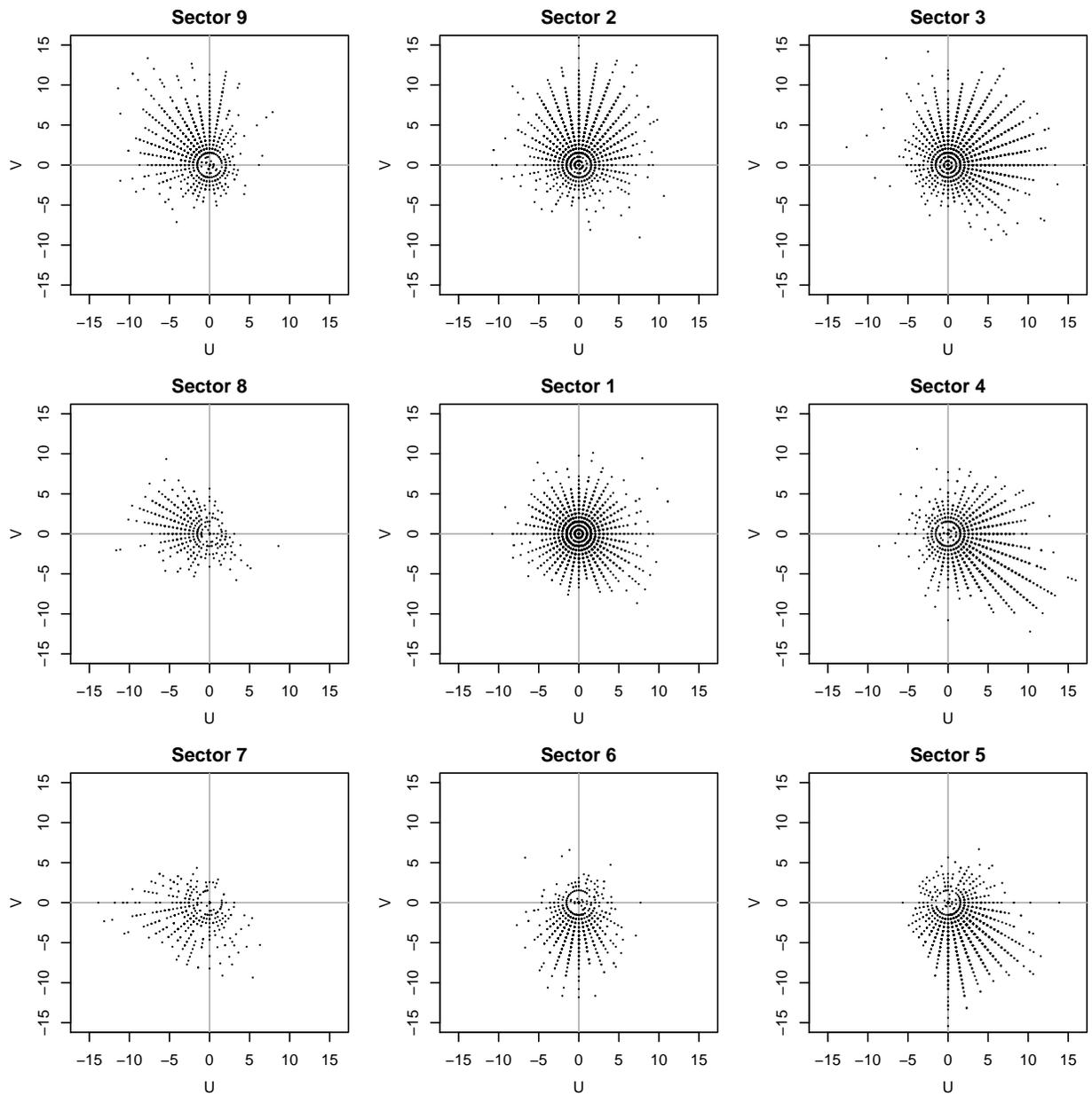}
\caption{Wind vector observations over the Pacific Northwest in 2007,
  conditional on the ensemble mean forecast falling into one of the
  sectors defined in Figure \ref{fig:sections}.  The unit for the wind
  components is meters per second.  \label{fig:scatterplots}}
\end{figure}

To motivate our specification of $\rho_{uv}$, we consider ensemble
forecast and observation data from the UWME in calendar year 2007,
comprising a total of 23,250 forecasts cases at 79 meteorological
stations in the Pacific Northwest.  Figure~\ref{fig:sections}
distinguishes nine sectors for $(u,v)$ wind vector forecasts and
observations.  Figure~\ref{fig:scatterplots} shows wind vector
observations in 2007, conditionally on the ensemble mean wind vector
falling into any given sector.  The conditional distributions are
partly elliptically contoured, particularly in the first sector, and
partly skewed, with an orientation that depends strongly on the
sector, and they reflect the discretized nature of wind observations,
as discussed in more detail in Section~\ref{sec:case.study}.

\begin{figure}[t]
\centering
\includegraphics[width=\textwidth]{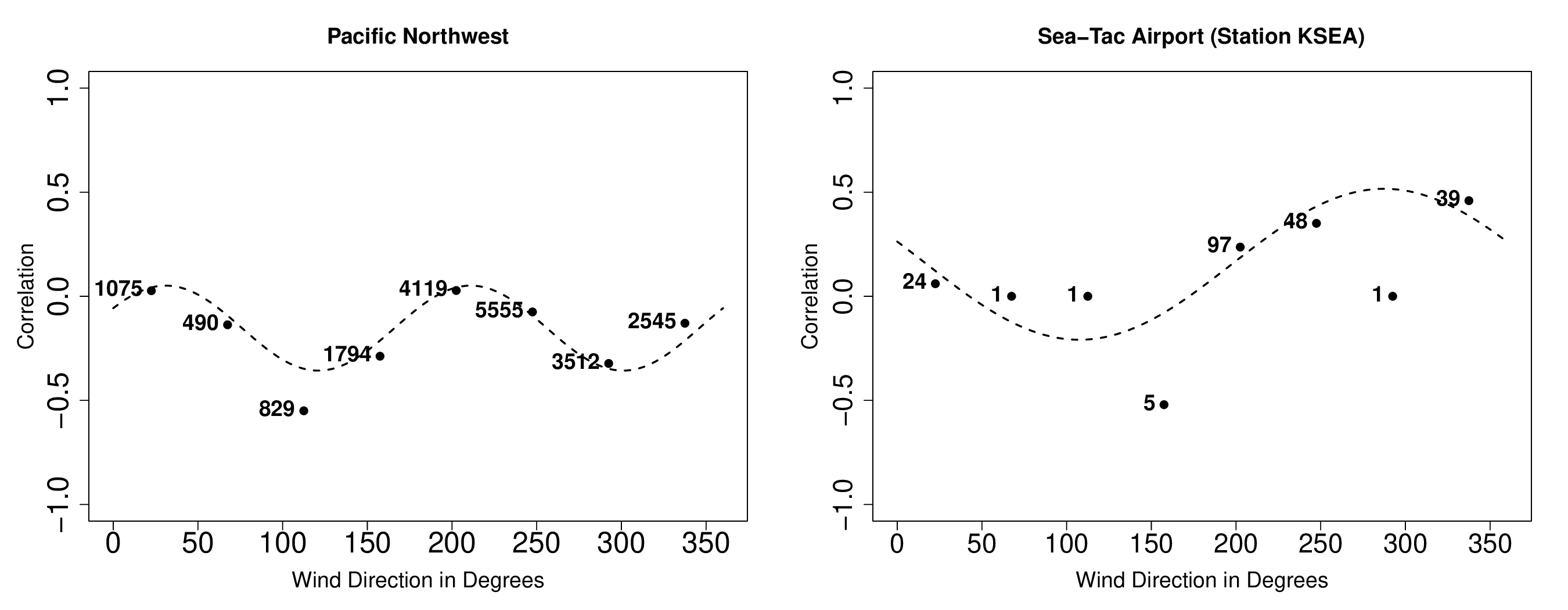}
\caption{Left: Correlation coefficient between the wind components in
  the scatter plots in Figure \ref{fig:scatterplots} as a function of
  the wind direction that corresponds to the center of the sector.
  The correlation coefficients and observation counts correspond to
  sectors 2--9, respectively.  The dashed curve shows the correlation
  model (\ref{eq:corr}) as fitted by the weighted least squares
  method.  Right: Same as left, but considering data from Sea-Tac
  Airport only.  \label{fig:corr}}
\end{figure}

The left-hand panel in Figure~\ref{fig:corr} plots the correlation
coefficient between the wind components in the scatter plots in Figure
\ref{fig:scatterplots} as a function of the wind directions that
correspond to the centers of sectors 2--9.  The panel demonstrates
that the ensemble mean wind direction ought to have a profound
influence on the correlation coefficient $\rho_{uv}$ in the
postprocessed EMOS density forecast (\ref{eq:BVN}).  Thus, we model
the correlation coefficient $\rho_{uv}$ as a trigonometric function of
the ensemble mean wind direction, $\theta$, measured in degrees, in
that
\begin{equation}  \label{eq:corr} 
\rho_{uv} = r 
\cos \! \left( \frac{2\pi}{360} \left( k \hsp \theta + \varphi \right) \! \right) + s,
\end{equation} 
where the parameters $r$, $s$, $k$ and $\varphi$ are estimated from
training data, in ways described below.  The coefficients $r$ and $s$
concern the overall magnitude of the correlation coefficient and need
to satisfy $|r| + |s| \leq 1$.  The parameter $k$ corresponds to the
number of periods of the trigonometric function, which we constrain to
be either $k = 1$, 2 or 3, and the direction $\varphi$ encodes phase
information.

We apply the correlation model (\ref{eq:corr}) to forecasts with
ensemble mean vectors in all nine wind sectors, including the first
sector.  Alternatively, if the ensemble mean wind vector falls into
the central first sector, one can take $\rho_{uv}$ to be equal to the
empirical correlation coefficient in the corresponding scatterplot.
In our experiences with the UWME, the two approaches resulted in
nearly identical predictive performance.

\subsection{Estimation}\label{sec:estimation}

Our EMOS postprocessed density forecast for a wind vector takes the
form of the bivariate normal probability density (\ref{eq:BVN}), where
the relationships of $\mu_u, \mu_v, \sigma_u^2, \sigma_v^2$ and
$\rho_{uv}$ to the raw ensemble forecast are specified in equations
(\ref{eq:mean}), (\ref{eq:variance}) and (\ref{eq:corr}).  It remains
to estimate the parameters that govern these equations, and we address
this task in three phases.

Before describing the phases of our estimation scheme, it is worth
noting that we follow \citet{Thorarinsdottir2010} and consider two
distinct variants, which we call the Regional EMOS and the Local EMOS
method, respectively.  In the Regional EMOS method, only one set of
parameters is estimated and used to produce forecasts over the entire
ensemble domain, such as the Pacific Northwest for the UWME.  Training
sets thus comprise data from all stations.  In contrast, the Local
EMOS method uses training data from the station at hand only, and thus
obtains a distinct set of parameters for each station.

We now turn to the first phase of our estimation scheme, in which we
fit the correlation model (\ref{eq:corr}).  We do this offline, once
and for all, based on historic, out-of-sample forecast and observation
data.  Specifically, for the UWME, we use data from calendar year 2007
to form the conditional scatterplots in Figure \ref{fig:scatterplots}
and compute and plot the corresponding empirical correlation
coefficients, as illustrated in Figure~\ref{fig:corr}.  We then decide
about a suitable value for the number of cycles, $k$, and fit the
remaining parameters, $r$, $s$ and $\varphi$, of the correlation model
(\ref{eq:corr}) by a weighted non-linear least squares technique,
using the {\sc R} function {\sc nls} with the weights being
proportional to the number of observations in the sectors
\citep{R2011}.  The use of a weighted least squares technique is
critical, particularly for the Local EMOS technique, as local wind
patterns may result in very few observations being available in any
given sector.  This first phase of the estimation is done once and for
all, using historic data from 2007, and the fitted correlation model
is applied throughout calendar year 2008, which we took as our test
period.

The left-hand panel in Figure~\ref{fig:corr} shows the fitted
correlation model for the Regional EMOS method, where we chose $k = 2$
and obtained weighted least squares estimates of $r = 0.20$, $s =
-0.15$ and $\varphi = - 61.9$ degrees, respectively.  As noted, the
Regional EMOS method uses these parameters throughout the UWME domain
and throughout the test period in calendar year 2008.  The right-hand
panel shows the Local EMOS correlation model at Seattle-Tacoma
(Sea-Tac) Airport, where we took $k = 1$ and obtained weighted least
squares estimates of $r = 0.24$, $s = 0.07$ and $\varphi = 70.5$
degrees, respectively.  The fitted Local EMOS correlation models at
the remaining stations in the UWME domain are provided and illustrated
in the appendix of \citet{Schuhen2011}.

In contrast to the first phase of our estimation scheme, the second
and third stages proceed on-line, that is, they use rolling training
periods consisting of data from the recent past.  The Regional EMOS
method uses all available data from the Pacific Northwest from the
last $n$ days prior to the forecast being made.  Missing data are
simply omitted from the training set.  For the Local EMOS method, the
training period comprises data from the station at hand from the $n$
most recent days where forecasts and observations were available.  In
either case, we talk of a sliding $n$-day training period.

In the second phase of our estimation scheme, the parameters $a_u$,
$b_u$, $a_v$ and $b_v$ in the specification (\ref{eq:mean}) of the
mean vector are estimated from the training data by standard linear
least squares regression.  In the third phase, the parameters $c_u$,
$d_u$, $c_v$ and $d_v$ in the specification (\ref{eq:variance}) of the
marginal variances are estimated on the same rolling training period
by the maximum likelihood technique, with all other parameters being
held fixed.  In other words, we maximize the logarithm of the
likelihood function, namely
\begin{equation}  \label{eq:log.lik} 
l(c_u,d_u,c_v,d_v) = {\textstyle \sum_{(x,t)}} \hhsp \log f^{(x,t)}(c_u,d_u,c_v,d_v),    
\end{equation} 
as a function of the parameters $c_u$, $d_u$, $c_v$ and $d_v$ in the
variance model (\ref{eq:variance}), which are all constrained to be
nonnegative.  The sum in the log likelihood function extends over all
locations $x$ and times $t$ for which there are data in the training
set.  Any single term of the form $f^{(x,t)}(c_u,d_u,c_v,d_v)$ refers
to the bivariate normal density (\ref{eq:BVN}) evaluated at the
verifying values $u = u^{(x,t)}$ and $v = v^{(x,t)}$, with $\mu_u$,
$\mu_v$ and $\rho_{uv}$ set at the numerical values implied by the
mean model (\ref{eq:mean}) and the correlation model (\ref{eq:corr}),
based on the parameter estimates from the first two phases of our
estimation scheme, and putting $\sigma_u^2 = c_u + d_u s_u^{2 \mid
  (x,t)}$ and $\sigma_v^2 = c_v + d_v s_v^{2 \mid (x,t)}$, where
$s_u^{2 \mid (x,t)}$ and $s_v^{2 \mid (x,t)}$ denote the ensemble
variances at location $x$ and valid time $t$, respectively.  The
optimization is performed numerically with the {\sc optim} function in
{\sc R}, using the Broyden-Fletcher-Goldfarb-Shanno algorithm with
initial values provided by the previous day's estimates.

The interpretation of the right-hand side of (\ref{eq:log.lik}) as a
log likelihood function is valid only if the forecast errors are
independent between times and locations.  While this is usually not
the case, an alternative interpretation as a mean logarithmic score
permits us to view the estimates as optimum score estimates, which are
tailored to the estimation of forecasting models \citep{Gneiting2005}.

\subsection{Example}  \label{sec:example}

\begin{figure}[p]
\centering
\includegraphics[width=\textwidth]{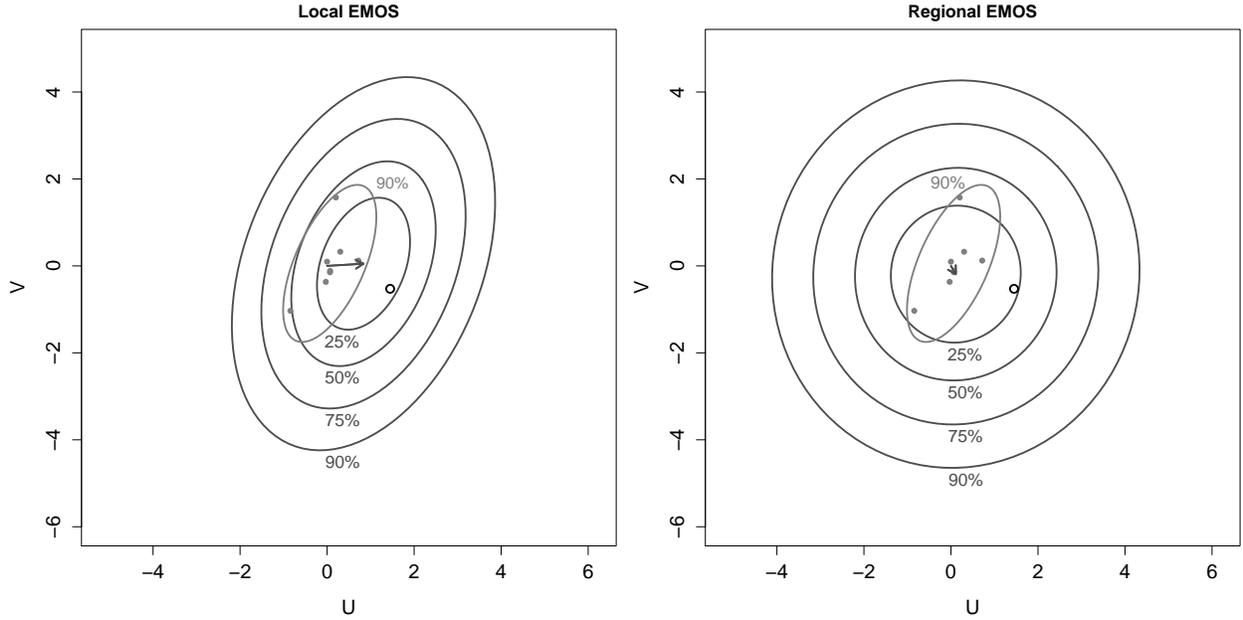}
\caption{Contour plot of the postprocessed Local EMOS (left) and
  Regional EMOS (right) density forecasts, along with the raw ensemble
  forecast, of the surface wind vector at Sea-Tac Airport, valid
  October 20, 2008 at 00 UTC, at a prediction horizon of 48 hours.
  The eight members of the University of Washington Mesoscale Ensemble
  (UWME; Eckel and Mass 2005) are shown as gray dots, along with the
  90\% prediction ellipse, which is based on a bivariate Gaussian fit
  to the ensemble values.  The 25\%, 50\%, 75\% and 90\% prediction
  ellipses and the mean vector for the postprocessed EMOS density
  forecast are shown in dark gray.  The verifying wind vector is
  represented by the small black circle at $(u,v) = (1.45, -0.53)$.
  The units are in meters per second.  \label{fig:example}}
\end{figure}

Figure \ref{fig:example} and Table \ref{tab:example} illustrate the
postprocessed Local and Regional EMOS density forecasts of the surface
wind vector at Sea-Tac Airport, valid October 20, 2008 at 00 UTC, at a
prediction horizon of 48 hours.  Thus, the forecast concerns the same
valid time and the same prediction horizon as the Local EMOS forecasts
illustrated in Figure \ref{fig:PugetSound}, where the station at
Sea-Tac Airport is located in the south-east Puget Sound area.  The
postprocessed density forecasts correct for the biases and
underdispersion in the raw UWME.  The Local EMOS forecast retains the
positive correlation structure in the raw ensemble and is sharper than
the Regional EMOS forecast.

\begin{table}[p]
\centering
\caption{Predictive means, variances and correlation for the
  postprocessed Local and Regional EMOS density forecasts in Figure
  \ref{fig:example} of the surface wind vector at Sea-Tac Airport,
  valid October 20, 2008 at 00 UTC, at a prediction horizon of 48
  hours.  The parameters refer to the general bivariate normal
  probability density (\ref{eq:BVN}), with the wind components
  represented in meters per second.  \label{tab:example}}
\begin{tabular}{lcc}
\toprule 
Method & Local EMOS & Regional EMOS \\
\midrule
$\mu_u$      & 0.84 & $\hphantom{-}$0.11 \rule{0mm}{5mm} \\
$\mu_v$      & 0.05 &            $-$0.19 \rule{0mm}{5mm} \\
$\sigma_u^2$ & 1.99 & $\hphantom{-}$3.87 \rule{0mm}{5mm} \\
$\sigma_v^2$ & 4.00 & $\hphantom{-}$4.31 \rule{0mm}{5mm} \\
$\rho_{uv}$  & 0.33 &            $-$0.02 \rule{0mm}{5mm} \\
\bottomrule
\end{tabular}
\end{table}

\section{Case study: Forecasting surface wind vectors over the Pacific Northwest}  
\label{sec:case.study}

We now consider the out-of-sample predictive performance of our
two-dimensional EMOS technique in a case study for wind vector
forecasts over the North American Pacific Northwest in 2008, based on
the University of Washington Mesoscale Ensemble (UWME; Eckel and Mass
2005).  We compare to the raw ensemble forecast and various reference
techniques, such as ensemble copula coupling, and assess the
performance of the bivariate EMOS technique when forecasts of wind
speed are desired only.

\subsection{University of Washington Mesoscale Ensemble}  \label{sec:UWME}

In our case study, the test set consists of forecasts of surface (10
meter) wind vectors based on the University of Washington Mesoscale
Ensemble (UWME; Eckel and Mass 2005) with valid date in calendar year
2008, at a prediction horizon of 48 hours.  The UWME is an
eight-member multi-analysis ensemble then based on the Fifth-Generation
Penn State/NCAR Mesoscale Model (MM5) with initial and lateral
boundary conditions obtained from operational centers around the
world.  The forecasts are made on a 12 km grid and the region covered
is the Pacific Northwest region of Western North America, including
the US states of Washington, Oregon and Idaho, as well as the southern
part of the Canadian province of British Columbia.  The forecasts were
bilinearly interpolated from the four surrounding grid points to the
observation locations and rotated to match the true direction at each
station.

Surface wind vector observations were provided by the weather
observation stations in the Automated Surface Observing System network
\citep{Service1998}.  The vector wind quantity studied here is
horizontal instantaneous surface wind, where `instantaneous' means
that the wind was measured and averaged over the last two minutes
before the valid time at 00 UTC.  The wind vector observations were
recorded as wind speed and wind direction, where wind speed was
rounded to the nearest whole knot, where a knot equals 0.5144 meters
per second, while values below two knots were recorded as zero.  The
observations are thus discretized, as is easily recognizable in Figure
\ref{fig:scatterplots}.  Quality control procedures as described in
\citet{Baars2005} were applied to the entire data set, removing dates
and locations with any missing forecasts or observations.

For calendar year 2008, 19,282 pairs of ensemble forecasts and
observations were available on 291 distinct days and at 79 distinct
observation locations.  Additional data from the years 2006 and 2007
were used to provide an appropriate rolling training period for all
days in 2008, for the first phase estimation of the correlation model
(\ref{eq:corr}) and to establish the optimal length of the rolling
training period.  Further information about the UWME, now using the
WRF mesoscale model, as well as real time forecasts and observations,
can be found online at
\url{http://www.atmos.washington} \url{.edu/~ens/uwme.cgi}.

\subsection{Training periods for Regional and Local EMOS}  \label{sec:regional.local}

As noted, we distinguish Regional and Local EMOS forecasts.  The
Regional EMOS method uses all available training data to estimate a
single set of parameters that is used throughout the Pacific Northwest
domain, and can be used directly on the model grid as well.  The Local
EMOS technique uses training data from the station at hand only to
obtain a distinct set of parameters at each station.  Thus, the
method applies at observation stations only, and can not be used
directly on the model grid.

\begin{figure}[t]
\centering
\includegraphics[width=\textwidth]{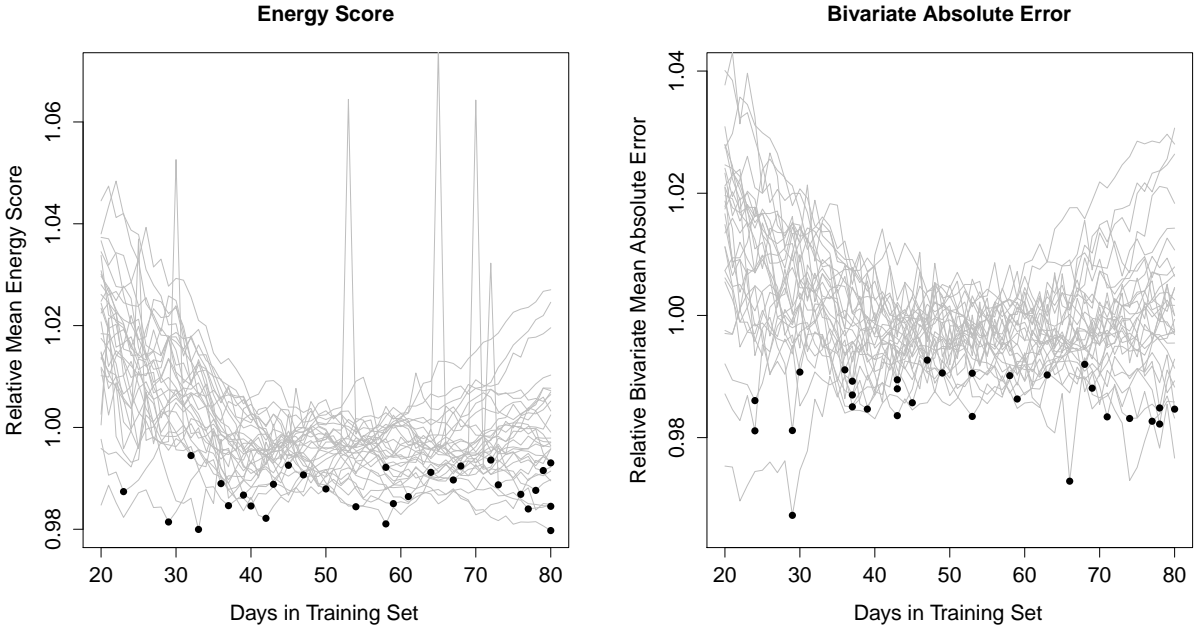}
\caption{Predictive performance of the Local EMOS method in calendar
  year 2007 as a function of the length of the rolling training period
  in terms of the relative mean energy score (left) and the relative bivariate mean
  absolute error (right).  Each curve corresponds to an observation
  station in Washington state, with the dot indicating the lowest
  value of the performance measure, relative to the average value 
  among the lengths considered.  \label{fig:training}}
\end{figure}

To determine a suitable value of the length $n$ of the rolling
training period for phases two and three of our estimation scheme, as
described in Section \ref{sec:EMOS}, 
we considered experiments
with wind vector forecasts based on the UWME in calendar year 2007.
In these experiments, phase one of the estimation scheme used (here,
in-sample) data from 2007, while phases two and three were done on a
rolling training period of length $n$ days.  The predictive
performance was evaluated using the mean energy score and the
bivariate mean absolute error, as described in the Appendix.  For the
Regional EMOS method, these metrics differed by less than a half
percent as the length $n$ of the rolling training period varied between
25 and 60.  Figure \ref{fig:training} summarizes the results for the
Local EMOS method, where we follow \citet{Thorarinsdottir2010} and
consider the observational locations in Washington state.  Based on
these results, our case study in calendar 2008 uses a rolling training
period of length $n = 30$ days for the Regional EMOS method, and of
length $n = 40$ days for the Local EMOS technique.  However, training
periods of any length between $n = 20$ and $n = 80$ days work well,
and the predictive performance of the EMOS methods is insensitive to
this choice.

\subsection{Reference forecasts}  \label{sec:reference}

We compare the postprocessed Regional and Local EMOS forecasts to the
raw UWME forecast, as well as to postprocessed reference forecasts, as
described now.  For all reference forecasts, we distinguish Regional
and Local methods, using rolling training periods of the most recent
available $n = 30$ and $n = 40$ days, respectively.

A natural reference standard is the {\bf Independent EMOS} technique,
which applies the standard EMOS or heterogeneous Gaussian regression
technique of \citet{Gneiting2005} to each of the vector wind
components $u$ and $v$ individually, and then combines them under the
assumption of independence.  This results in a postprocessed bivariate
normal density forecast of the form (\ref{eq:BVN}) with the critical
correlation parameter $\rho_{uv}$ constrained to be zero, but with the
means and variances for $u$ and $v$ being essentially identical to
those in the bivariate EMOS method, except for minor differences due
to the slightly differing estimation schemes.  In particular, the
bivariate EMOS method and the Independent EMOS technique yield
essentially identical deterministic, bivariate mean and median wind
vector forecasts, even though the respective bivariate predictive
densities may differ substantially.

The {\bf Ensemble Copula Coupling (ECC)} method, originally hinted at
by \citet{Bremnes2007} and \citet{Krzysztofowicz2008} and recently
investigated and developed by \cite{Schefzik2011}, is a tool for
restoring a raw ensemble's flow-dependent rank dependence structure in
individually postprocessed predictive distributions.  Here, we
describe and apply the method in the context of the Independent EMOS
technique and a raw ensemble with $m$ members.  At any given location
and prediction horizon, let $u_1^*, \ldots, u_m^*$ and $v_1^*, \ldots,
v_m^*$ denote samples from the individually postprocessed univariate
EMOS distributions.  As the postprocessed distributions for the wind
vector components are univariate Gaussian, these are simply random
numbers drawn from a univariate normal density.  Let $u_1, \ldots,
u_m$ and $v_1, \ldots, v_m$ denote the raw ensemble values for the
respective wind components, and let $\tau_u$ and $\tau_v$ be
permutations of the numbers $1, \ldots, m$ such that
\[
u_{\tau_u(1)} \leq u_{\tau_u(2)} \leq \cdots \leq u_{\tau_u(m)} 
\quad \mbox{and} \quad
v_{\tau_v(1)} \leq v_{\tau_v(2)} \leq \cdots \leq v_{\tau_v(m)}, 
\] 
with any ties resolved at random.  The ECC ensemble then consists of
the $m$ wind vectors
\[
(u_{\tau_u(1)}^*, v_{\tau_v(1)}^*), \ldots, (u_{\tau_u(m)}^*, v_{\tau_v(m)}^*).
\]
Thus, the ECC ensemble inherits and honors the raw ensemble's rank
dependence structure.  For example, if the first raw ensemble member
shows the second lowest $u$ component and the third highest $v$
component among the raw ensemble members, then the same property holds
true for the ECC ensemble.

\citet{Schefzik2011} provides a detailed discussion of the ECC
technique and exposes its ties to copulas \citep{Schoelzel2008}.  The
wind vector postprocessing technique recently proposed by
\citet{Pinson2011} can be interpreted as a particularly attractive
variant, where the values $u_1^*, \ldots, u_m^*$ are constructed as a
translation and dilation of the raw ensemble values $u_1, \ldots,
u_m$, and $v_1^*, \ldots, v_m^*$ as a translation and dilation of
$v_1, \ldots, v_m$, respectively, rather than being drawn at random.
Consequently, the postprocessed ensemble forecast retains both the raw
ensemble's bivariate Spearman rank correlation and its (standard)
Pearson product moment correlation structure.

Finally, we consider an {\bf Error Dressing} ensemble as proposed by
\citet{Gneiting2008a} in the spirit of the work of
\citet{Roulston2003}, where we dress the UWME mean forecast with 35
error vectors from the corresponding training set.

\subsection{Results over the Pacific Northwest}  \label{sec:PNW}

We now give verification results aggregated over our test period, the
calendar year 2008, and the Pacific Northwest domain of the UWME.  For
details about the verification techniques method, which include the
multivariate rank histogram, as proposed by \citet{Gneiting2008a} for
calibration checks, and proper scoring rules \citep{Gneiting2007}, are
given in the Appendix.  All scores used are negatively oriented, that
is, the smaller the better.

\begin{figure}[p]
\centering
\includegraphics[width=0.575\textwidth]{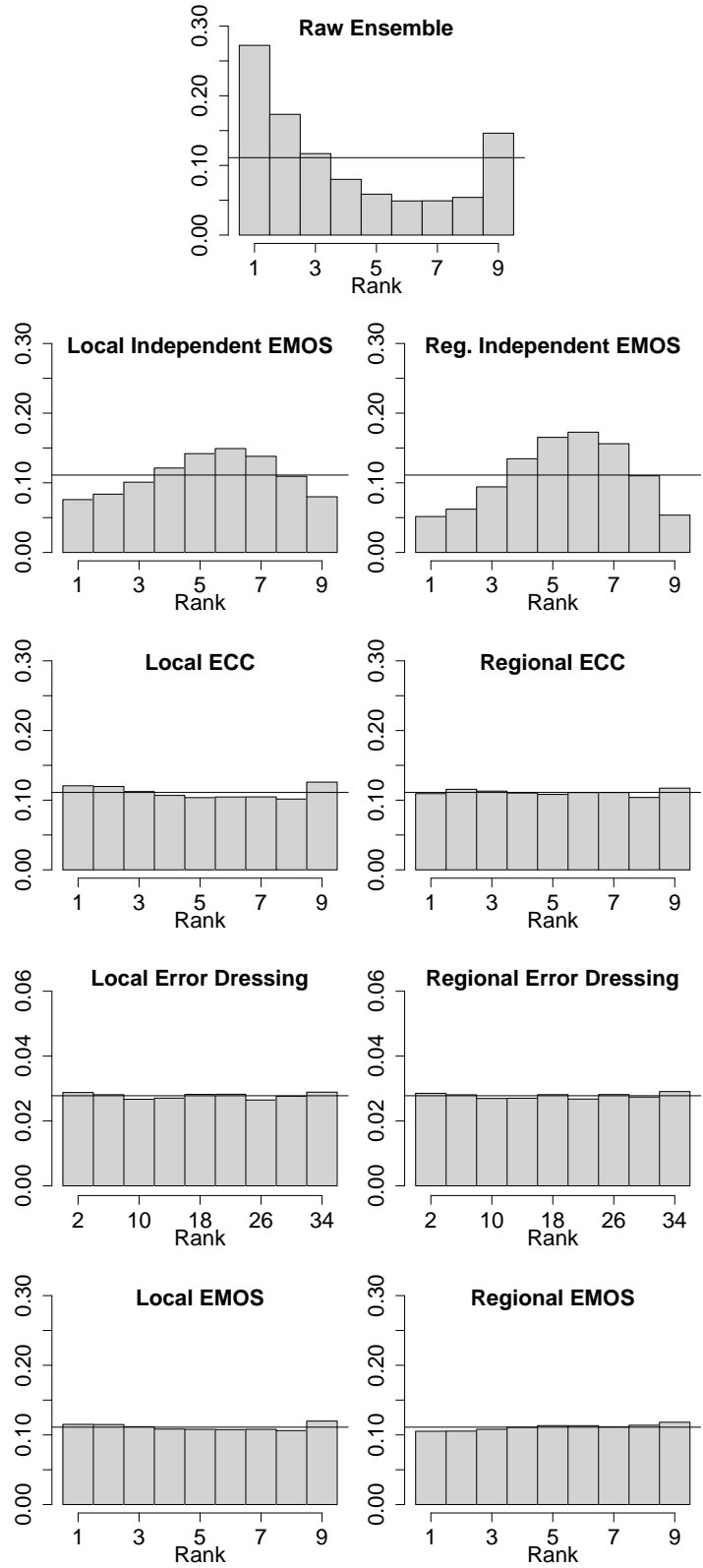}
\caption{Multivariate rank histograms for raw and postprocessed
  ensemble forecasts of surface wind vectors, aggregated over calendar
  year 2008 and the Pacific Northwest.
  \label{fig:MRH}}
\end{figure}

Figure \ref{fig:MRH} shows multivariate rank histograms for the raw
ensemble forecast and Regional and Local versions of the Independent
EMOS, ECC, Error Dressing and (bivariate) EMOS techniques.  The raw
ensemble forecast shows a U-shaped rank histogram, thus indicating
underdispersion.  In contrast, the Independent EMOS forecasts are
overdispersed, as they neglect to take dependencies between the wind
vector components into account, which can be corrected for with the
ECC technique.  In addition to the ECC method, the Error Dressing and
(bivariate) EMOS techniques show uniform rank histograms, as expected
from a calibrated forecast.

\begin{table}[t]
\centering 
\caption{Predictive performance of forecasts of surface wind vectors
  in terms of the mean energy score (ES) and the bivariate mean
  absolute error (bMAE), both in meters per second, and the
  reliability index $\Delta$ for the multivariate rank histogram,
  aggregated over calendar year 2008 and the Pacific
  Northwest.  \label{tab:PNW}}
\begin{tabular}{lccc}
\toprule 
Method & ES & bMAE & $\Delta$ \\
\midrule
Raw Ensemble              & 2.47 & 3.01 & 0.53 \\
\midrule
Regional Independent EMOS & 2.43 & 2.79 & 0.37 \\
Regional ECC              & 2.33 & 3.00 & 0.02 \\
Regional Error Dressing   & 2.19 & 3.01 & 0.01 \\
Regional EMOS             & 2.01 & 2.80 & 0.03 \\
\midrule
Local Independent EMOS    & 2.28 & 2.60 & 0.21 \\
Local ECC                 & 2.16 & 2.78 & 0.07 \\
Local Error Dressing      & 2.07 & 2.87 & 0.01 \\
Local EMOS                & 1.87 & 2.61 & 0.03 \\
\bottomrule
\end{tabular}
\medskip
\end{table}

Table~\ref{tab:PNW} provides numerical summary measures of the
predictive performance, with the reliability index $\Delta$ for the
multivariate rank histogram confirming the visual impression in Figure
\ref{fig:MRH}.  The energy score is a direct analogue of the
continuous ranked probability score for univariate quantities that
provides an overall assessment of the quality of a probabilistic
forecast, addressing both calibration and sharpness.  The bivariate
absolute error generalizes the absolute error and assesses
deterministic forecast skill.

Not surprisingly, the Local approaches outperform the Regional
approaches, both in terms of the mean energy score and the bivariate
mean absolute error.  Also, the postprocessed Independent EMOS
forecasts show lower scores than the raw forecast.  The Independent
EMOS and the (bivariate) EMOS forecasts have nearly identical
bivariate mean absolute error, at a substantially lower value than for
the other types of forecasts, including the ECC approach.  This effect
can be attributed to a discretization effect, in that the ECC
technique turns the Independent EMOS density forecast into a discrete
ensemble forecast.  However, the ECC approach improves on the
Independent EMOS forecast in terms of the energy score, as it honors
the raw ensemble's flow-dependent bivariate dependence structure.  The
Error Dressing technique also shows good probabilistic forecast skill,
as evidenced by a low energy score, even though it is unable to match
the scores of the (bivariate) EMOS method, which performs the best.
When compared to the raw ensemble, the postprocessed Local EMOS
forecast reduces the mean energy score from 2.47 to 1.87 meters per
second.

\subsection{Results at Sea-Tac Airport}  \label{sec:KSEA} 

Next we consider forecasts at the observation station at Sea-Tac
Airport, with summary measures of the predictive performance in
calendar year 2008 being provided in Table~\ref{tab:KSEA}.  The
results mimic those for the Pacific Northwest.  The Local EMOS
forecast shows the highest probabilistic forecast skill, as quantified
by the energy score, which decreases to 1.94 meters per second, as
compared to 2.25 meters per second for the raw ensemble.

\begin{table}[p]
\centering
\caption{Predictive performance of forecasts of surface wind vectors
  at Sea-Tac Airport in calendar year 2008 in terms of the mean energy
  score (ES) and the bivariate mean absolute error (bMAE), both in
  meters per second, and the reliability index $\Delta$ for the
  multivariate rank histogram.  \label{tab:KSEA}}
\begin{tabular}{lccc}
\toprule 
Method & ES & bMAE & $\Delta$ \\
\midrule
Raw Ensemble              & 2.25 & 2.77 & 0.54 \\
\hline
Regional Independent EMOS & 2.41 & 2.89 & 0.29 \\
Regional ECC              & 2.37 & 3.06 & 0.23 \\
Regional Error Dressing   & 2.12 & 2.92 & 0.07 \\
Regional EMOS             & 2.06 & 2.90 & 0.24 \\
\hline
Local Independent EMOS    & 2.40 & 2.73 & 0.22 \\
Local ECC                 & 2.21 & 2.87 & 0.22 \\
Local Error Dressing      & 1.98 & 2.75 & 0.02 \\
Local EMOS                & 1.94 & 2.74 & 0.09 \\
\bottomrule
\end{tabular}
\medskip
\end{table}

\begin{figure}[p]
\centering
\includegraphics[width=0.675\textwidth]{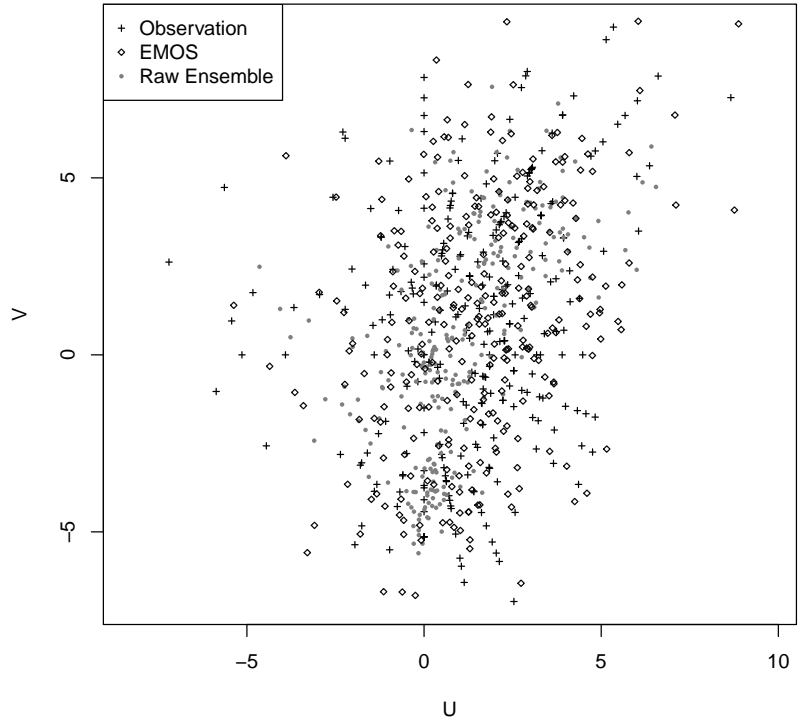}
\caption{Marginal calibration diagram for forecasts of surface wind
  vectors at Sea-Tac Airport in calendar year 2008.  For each day
  available, the plot shows the observed wind vector, a randomly
  chosen member of the raw ensemble, and a wind vector sampled from
  the postprocessed Local EMOS density
  forecast.  \label{fig:marg.cal}}
\end{figure}

For an illustration and explanation of how and to what extent the
Local EMOS technique succeeds in improving the raw ensemble forecast
at Sea-Tac Airport, consider the display in Figure \ref{fig:marg.cal},
which is a bivariate variant of the marginal calibration diagram
proposed by \citet{Gneiting2007a}.  Essentially, a marginal calibration
diagram compares the observed climatology to the climatology incurred
by the forecasts.  In our display we plot, for each day for which data
are available, the observed wind vector, perturbed very slightly in
order to honor the undiscretized distribution and improve readability,
a randomly chosen member of the UWME raw ensemble, and a wind vector
sampled from the postprocessed, bivariate normal Local EMOS density
forecast.  Again we observe that the raw ensemble is underdispersive
and fails to predict any extreme wind vectors.  The postprocessed
Local EMOS forecast corrects for the underdispersion, thus leading to
substantially improved probabilistic forecast skill, as evidenced by
the energy score.  However, the raw ensemble does not show any
recognizable biases, and so the improvement in deterministic forecast
skill is minor.

\subsection{Results for wind speed}  \label{sec:speed}

In addition to producing calibrated and sharp forecasts of wind
vectors, the bivariate EMOS method can be used to predict wind speeds,
and so can be compared to the technique of Thorarinsdottir and Gneiting
(2010), which is custom tailored to this task.
As wind speed is a nonnegative quantity, we employ truncated normal
predictive distributions, using an estimation scheme that is based on
optimum score estimation, both in Regional and Local versions, where
we use rolling training periods comprising the most recent $n = 30$
and $n = 40$ days available, respectively.  In what follows, we refer
to the method of \citet{Thorarinsdottir2010} as wind speed EMOS.

To generate probabilistic forecasts of wind speed from the bivariate
EMOS forecast, we sample one hundred wind vectors from the bivariate
predictive distribution, and compute the Euclidean norm of each
vector, thereby obtaining a discrete forecast ensemble of size $m =
100$ for wind speed.  Table~\ref{tab:speed} compares the predictive
performance of this approach to that of the wind speed EMOS technique.
As noted, the wind speed observations are strongly discrete, with wind
speeds below two knots recorded as zero, which applies to about 14\%
of the observations in the test period.  From the perspective of wind
vectors, these observations tend to fall into the center of the
respective predictive distribution, and so the effect of the
discretization is weak.  From the perspective of wind speeds, an
observed value of zero is right at the boundary of the climatological
range, and so the effect is nonnegligible.  To account for it, we
replace every observation of zero by a number drawn uniformly and at
random between zero and two knots, or between zero and 1.03 meters per
seconds, when computing the corresponding continuous ranked
probability score or absolute error.  The bivariate EMOS technique,
particularly in its Local version, nearly matches the predictive
performance of the specialized wind speed EMOS technique of
\citet{Thorarinsdottir2010}.

\begin{table}[t]
\centering
\caption{Predictive performance of forecasts of wind speed in terms of
  the mean continuous ranked probability score (CRPS) and the mean
  absolute error (MAE), in meters per second, averaged over calendar
  year 2008 and the Pacific Northwest.  \label{tab:speed}} 
\begin{tabular}{lcc}
\toprule 
Method & CRPS & MAE \\
\midrule
Raw Ensemble             & 1.34 & 1.68 \\
\midrule
Regional Wind Speed EMOS & 1.11 & 1.57 \\
Regional EMOS            & 1.15 & 1.63 \\
\midrule
Local Wind Speed EMOS    & 1.08 & 1.50 \\
Local EMOS               & 1.07 & 1.51 \\
\bottomrule
\end{tabular}
\medskip
\end{table}

\section{Discussion}  \label{sec:discussion}

In this paper, we have proposed a bivariate EMOS approach to the
statistical postprocessing of ensemble forecasts of wind vectors that
results in bivariate normal density forecasts.  In experiments with
48-hour ahead forecasts of surface wind vectors over the Pacific
Northwest, based on the University of Washington Mesoscale Ensemble
(UWME), the postprocessed EMOS density forecast proved to bias correct
and calibrate the raw ensemble forecast, therefore resulting in
strongly improved deterministic and probabilistic predictive skill.
When compared to the raw ensemble forecast and aggregated over
calendar year 2008 and the Pacific Northwest, the Local EMOS technique
reduced the bivariate mean absolute error by 13\% and the bivariate
mean energy score by 24\%.

There are several directions into which our bivariate EMOS method
could be developed.  In phases two and three of the estimation scheme,
the exponential forgetting approach as in \citet{Pinson2011} could be
implemented, where the parameter estimates are updated in a
computationally efficient, adaptive way.  Furthermore, a
geostatistical approach such as that of \citet{Kleiber2011} in the
context of Bayesian model averaging \citep{Raftery2005} could be
developed in order to spread the Local EMOS estimates, which currently
are available at observation locations only, over the model grid.
Alternatively, if an analysis is used to fit the EMOS model, as was
done by \citet{Pinson2011}, training data are available on the
analysis grid, thereby allowing for a gridded Local EMOS approach.

A closer look at the scatterplots in Figure~\ref{fig:scatterplots}
suggests that the conditional distribution of the observed wind
vector, given the ensemble forecast, tends to be skewed.  Thus, our
bivariate EMOS approach, which currently is based on bivariate normal
densities, could be extended to allow for skewed distributions, such
as bivariate skew-normal or skew-$t$ densities, similar to the wind
vector time series methods proposed by \citet{HeringGenton2010}.
However, any such approach would be considerably more complex, and as
it is more difficult to estimate a more complex predictive model, it
is not clear whether or not such an extension would result in improved
forecast performance.

\citet{Sloughter2009} and \citet{Sloughter2011} proposed a bivariate
version of the Bayesian model averaging technique (BMA; Raftery et
al., 2005) for postprocessing ensemble forecasts of wind vectors.
\nocite{Raftery2005} Like our EMOS method, the BMA approach results in
a bivariate forecast density.  However, the BMA forecast density is a
finite mixture of bivariate, power-transformed normal densities and
thus can be multimodal, as opposed to the EMOS forecast density, which
is necessarily unimodal and elliptically symmetric.  Thus, the basic
setting of the EMOS approach is more parsimonious, thereby allowing
for the key innovation in our work, namely the explicit modeling of
the correlation between the $u$ and $v$ wind vector components,
conditionally on the direction of the ensemble mean vector.

In contrast to the BMA and EMOS methods, the ensemble copula coupling
technique (ECC; Schefzik 2011) and the postprocessing approach
proposed by \citet{Pinson2011} generate discrete forecast ensembles
that are constrained to have the same number of members, and the same
bivariate Spearman rank correlation coefficients, as the raw ensemble.
Therefore, these methods are particularly well adapted to large
ensembles, where there is no pronounced need for the transition from a
discrete forecast ensemble to a density forecast, nor any need for
statistical correction of the conditional correlation structure
between the wind vector components.  Accordingly, both
\citet{Pinson2011} and \citet{Schefzik2011} tailored their methods to
the 50-member European Centre for Medium-range Weather Forecasts
(ECMWF) ensemble, while our work considered the much smaller
eight-member UWME, where the wind vector ensemble forecasts for any
given location and valid time may show unreliable, physically
unrealistic empirical correlation coefficients.  This effect is caused
by the small ensemble size and corroborates the need for a parametric
correlation model.

In this paper, we considered probabilistic forecasts of wind vectors
for a single location and valid time.  While we modeled the bivariate
correlation structure between the wind vector components, we
considered the postprocessed bivariate density forecasts at each site
individually, without modeling the dependence structure between
locations.  To address the latter, the methods developed in our paper
could be combined with the spatial statistical techniques introduced
by \citet{Gel2004}.  A common advantage of the method proposed by
\citet{Pinson2011} and the ECC technique \citep{Schefzik2011} lies in
their immediate extension to the calibration of spatio-temporal
trajectories, in that the raw ensemble's bivariate rank dependence
structure is inherited by the postprocessed ensemble, subject to the
aforementioned caveats.

\section*{Appendix: Verification methods}  \label{Appendix} 

In verifying probabilistic forecasts of a multivariate weather
quantity, we use techniques introduced, studied and used by 
\citet{Gneiting2008a}, \citet{Gneiting2011} and \citet{PinsonHagedorn2011}.

To assess the calibration of probabilistic forecasts of wind vectors,
we use the multivariate rank histogram, which is a natural, direct
generalization of the verification rank histogram or Talagrand diagram
for a univariate quantity \citep{Anderson1996, Hamill1997,
  Talagrand1997} and can be interpreted analogously, in ways described
by \citet{Hamill2001}.  In particular, U-shaped multivariate rank
histograms correspond to underdispersed ensembles, while inverse U- or
hump-shaped histograms indicate overdispersed ensembles.  For a
calibrated ensemble, we expect a uniform rank histogram.  For an
ensemble with $m$ members, the multivariate verification rank is a
possibly randomized number between 1 and $m + 1$, and we refer to
\citet{Gneiting2008a} for the technical details in its construction.
To quantify the departure of the rank histogram from uniformity, we use
the discrepancy or reliability index $\Delta$ proposed by
\citet{DelleMonache2006}, given by
\begin{equation}  \label{eq:Delta} 
\Delta = \sum_{i=1}^{m+1} \left| f_i - \frac{1}{m+1} \right|, 
\end{equation}
where $f_i$ is the observed relative frequency of verification rank $i
= 1, \ldots, m + 1$.  The UWME raw ensemble and ensemble copula
coupling (ECC) techniques result in a discrete forecast ensemble with
$m = 8$ members, from which the construction of the multivariate rank
histogram is straightforward.  The Error Dressing techniques provide
discrete forecast ensembles with $m = 35$ members, and we bin the
corresponding 36 ranks in the multivariate rank histogram into nine
groups, comprising ranks 1--4, \ldots, 33--36, respectively, to
facilitate the comparison.  For the Independent EMOS and bivariate
EMOS techniques, we draw a simple random sample of size $m = 8$ from
the bivariate normal predictive distribution, and then compute the
multivariate rank histogram.

To assess the overall quality of probabilistic forecasts, considering
both calibration and sharpness, we use proper scoring rules
\citep{Gneiting2007, Wilks2011}.  For a univariate weather quantity,
such as wind speed, the proper continuous ranked probability score is
defined as
\begin{equation}  \label{eq:crps} 
{\rm crps}(P,y) 
  = \int_{-\infty}^\infty (P(x) - {\rm I}(x \geq y))^2 \: {\rm d}x 
  = \E_P |X-y| - \frac{1}{2} \hsp \E_P |X-X'|,
\end{equation}
where $P$ is the predictive distribution, here taking the form of a
cumulative distribution function, $X$ and $X'$ are independent random
variables with cumulative distribution function $P$, and $y$ is the
verifying value \citep{Gneiting2007}.  The term ${\rm I}(x \geq y)$
denotes an indicator function, equal to 1 if $x \geq y$, and equal to
0 otherwise, and $\E$ is the expectation operator.  The absolute error
is defined as
\begin{equation}  \label{eq:ae} 
{\rm ae}(P,y) = |{\rm med}_P - y|,
\end{equation}
where ${\rm med}_P$ is a median of the probability distribution $P$
\citep{Gneiting2011, PinsonHagedorn2011}.

The energy score was introduced by \citet{Gneiting2007} and
\citet{Gneiting2008a} as a direct generalization of the continuous
ranked probability score (\ref{eq:crps}) in the evaluation of
probabilistic forecasts of multivariate quantities.  As we are
interested in wind vectors, we restrict the discussion to bivariate
weather quantities.  The energy score then is defined as
\begin{equation}  \label{eq:es} 
{\rm es}(P,y) = \E_P \hhsp \| X - y \| - \frac{1}{2} \hsp \E_P \hhsp \| X - X' \|,
\end{equation} 
where $\| \cdot \|$ denotes the Euclidean norm in $\real^2$, $P$ is
the predictive distribution, $X$ and $X'$ are independent random
vectors with distribution $P$, and $y \in \real^2$ is the verifying
wind vector.  For an ensemble forecast, the predictive distribution
$P_{\rm ens}$ has point mass $\frac{1}{m}$ at the member forecasts
$x_1, \ldots, x_m \in \real^2$, and the energy score can be evaluated
as
\[
{\rm es}(P_{\rm ens},y) = \frac{1}{m} \sum_{j=1}^m \| x_i - y \|
- \frac{1}{2m^2} \sum_{i=1}^m \sum_{j=1}^m \| x_i - x_j \|.
\]
We use this formula to compute the energy score for the UWME raw
ensemble, Ensemble Copula Coupling (ECC) and Error Dressing
techniques.  For the Independent EMOS and bivariate EMOS techniques,
we draw a simple random sample $x_1, \ldots, x_k \in \real^2$ from the
corresponding predictive density, and replace the exact energy score
(\ref{eq:es}) by the computationally efficient approximation
\[
\widehat{\rm es}(P,y)
= \frac{1}{k} \sum_{i=1}^k \| x_i - y \| 
- \frac{1}{2 \hsp (k-1)} \sum_{i=1}^{k-1} \| x_i - x_{i+1} \|, 
\]
where we use a sample of size $k = 10,000$.  Similar approximations
apply to the continuous ranked probability score (\ref{eq:crps}).

The natural generalization of the absolute error (\ref{eq:ae}) is the
bivariate absolute error
\begin{equation}  \label{eq:bae} 
{\rm bae}(P,y) = \| {\rm bmed}_P - y \|, 
\end{equation}
where ${\rm bmed}_P$ denotes the bivariate or spatial median of the
probability distribution $P$, defined as
\[
{\rm bmed}_P = \arg {\textstyle \min_{\hsp x \hhsp \in \hhsp \real^2}} \, \E_P \| \hhsp x - X \|, 
\]
where $X$ is a random vector with distribution $P$ \citep{Vardi2000,
  Gneiting2011}.  For an elliptically symmetric distribution, such as
a bivariate normal distribution, the bivariate median and the mean
vector coincide.  For other types of bivariate distributions, such as
an ensemble forecast with point mass $\frac{1}{m}$ at the member
forecasts $x_1, \ldots, x_m \in \real^2$, the bivariate median
generally is different from the corresponding mean vector, and
typically it needs to be determined numerically.  For doing this we
use the algorithm described by \citet{Vardi2000} and implemented in
the {\sc R} package {\sc ICSNP}.
 
In practice, forecasting methods are assessed by averaging scores over
a test period, resulting in the mean continuous ranked probability
score (CRPS), mean absolute error (MAE), mean energy score (ES) and
bivariate mean absolute error (bMAE), respectively.  All these
quantities are negatively oriented, that is, the smaller the better,
and we report their values in the unit of meters per second.

\section*{Acknowledgements}

The authors are grateful to Jeff Baars, Tom M.~Hamill, Clifford
F.~Mass, Adrian E.~Raftery and J.~McLean Sloughter for discussions and
providing data.  Tom Hamill communicated to us the idea that underlies
ensemble copula coupling (ECC), well before we noticed the independent
discussions in the work of \citet{Bremnes2007} and
\citet{Krzysztofowicz2008} and the term ECC was coined in
\citet{Schefzik2011}.

\bibliography{windvector}

\begin{thebibliography}{}

\bibitem[\protect\citeauthoryear{Anderson}{Anderson}{1996}]{Anderson1996}
Anderson, J.~L. (1996).
\newblock {A method for producing and evaluating probabilistic forecasts from
  ensemble model integrations}.
\newblock {\em Journal of Climate\/}~{\em 9}, 1518--1530.

\bibitem[\protect\citeauthoryear{Baars}{Baars}{2005}]{Baars2005}
Baars, J. (2005).
\newblock {Observations QC documentation}.
\newblock Available at
  \url{http://www.atmos.washington.edu/mm5rt/qc_obs/qc_doc.html}.

\bibitem[\protect\citeauthoryear{Bao, Gneiting, Grimit, Guttorp, and
  Raftery}{Bao et~al.}{2010}]{Bao2010}
Bao, L., T.~Gneiting, E.~P. Grimit, P.~Guttorp, and A.~E. Raftery (2010).
\newblock {Bias correction and Bayesian model averaging for ensemble forecasts
  of surface wind direction}.
\newblock {\em Monthly Weather Review\/}~{\em 138}, 1811--1821.

\bibitem[\protect\citeauthoryear{Bremnes}{Bremnes}{2007}]{Bremnes2007}
Bremnes, J.~B. (2007).
\newblock {Improved calibration of precipitation forecasts using ensemble
  techniques. Part 2: Statistical calibration methods}.
\newblock {Norwegian Meteorological Institute, Technical Report no.~04/2007.
  Available at} \url{http://met.no/Forskning/Publi}
  \url{kasjoner/Publikasjoner_2007/filestore/report04_2007.pdf}.

\bibitem[\protect\citeauthoryear{{Delle Monache}, Hacker, Zhou, Deng, and
  Stull}{{Delle Monache} et~al.}{2006}]{DelleMonache2006}
{Delle Monache}, L., J.~P. Hacker, Y.~Zhou, X.~Deng, and R.~B. Stull (2006).
\newblock {Probabilistic aspects of meteorological and ozone regional ensemble
  forecasts}.
\newblock {\em Journal of Geophysical Research\/}~{\em 111}, D24307.

\bibitem[\protect\citeauthoryear{Eckel and Mass}{Eckel and
  Mass}{2005}]{Eckel2005}
Eckel, F.~A. and C.~F. Mass (2005).
\newblock {Aspects of effective mesoscale, short-range ensemble forecasting}.
\newblock {\em Weather and Forecasting\/}~{\em 20}, 328--350.

\bibitem[\protect\citeauthoryear{Gel, Raftery, and Gneiting}{Gel
  et~al.}{2004}]{Gel2004}
Gel, Y., A.~E. Raftery, and T.~Gneiting (2004).
\newblock Calibrated probabilistic mesoscale weather field forecasting: {T}he
  geostatistical output perturbation ({GOP}) method (with discussion).
\newblock {\em Journal of the American Statistical Association\/}~{\em 99},
  575--587.

\bibitem[\protect\citeauthoryear{Gneiting}{Gneiting}{2011}]{Gneiting2011}
Gneiting, T. (2011).
\newblock {Making and evaluating point forecasts}.
\newblock {\em Journal of the American Statistical Association\/}~{\em 106},
  746--762.

\bibitem[\protect\citeauthoryear{Gneiting, Balabdaoui, and Raftery}{Gneiting
  et~al.}{2007}]{Gneiting2007a}
Gneiting, T., F.~Balabdaoui, and A.~E. Raftery (2007).
\newblock {Probabilistic forecasts, calibration and sharpness}.
\newblock {\em Journal of the Royal Statistical Society Series B\/}~{\em 69},
  243--268.

\bibitem[\protect\citeauthoryear{Gneiting and Raftery}{Gneiting and
  Raftery}{2007}]{Gneiting2007}
Gneiting, T. and A.~E. Raftery (2007).
\newblock {Strictly proper scoring rules, prediction, and estimation}.
\newblock {\em Journal of the American Statistical Association\/}~{\em 102},
  359--378.

\bibitem[\protect\citeauthoryear{Gneiting, Raftery, Westveld, and
  Goldman}{Gneiting et~al.}{2005}]{Gneiting2005}
Gneiting, T., A.~E. Raftery, A.~H. Westveld, and T.~Goldman (2005).
\newblock {Calibrated probabilistic forecasting using ensemble model output
  statistics and minimum CRPS estimation}.
\newblock {\em Monthly Weather Review\/}~{\em 133}, 1098--1118.

\bibitem[\protect\citeauthoryear{Gneiting, Stanberry, Grimit, Held, and
  Johnson}{Gneiting et~al.}{2008}]{Gneiting2008a}
Gneiting, T., L.~I. Stanberry, E.~P. Grimit, L.~Held, and N.~A. Johnson (2008).
\newblock {Assessing probabilistic forecasts of multivariate quantities, with
  an application to ensemble predictions of surface winds}.
\newblock {\em Test\/}~{\em 17}, 211--235.

\bibitem[\protect\citeauthoryear{Grimit and Mass}{Grimit and
  Mass}{2002}]{Grimit2002}
Grimit, E. and C.~Mass (2002).
\newblock {Initial results of a mesoscale short-range ensemble forecasting
  system over the Pacific Northwest}.
\newblock {\em Weather and Forecasting\/}~{\em 17}, 192--205.

\bibitem[\protect\citeauthoryear{Hagedorn, Hamill, and Whitaker}{Hagedorn
  et~al.}{2008}]{Hagedorn2008}
Hagedorn, R., T.~M. Hamill, and J.~S. Whitaker (2008).
\newblock {Probabilistic forecast calibration using ECMWF and GFS ensemble
  reforecasts. Part I: Temperature}.
\newblock {\em Monthly Weather Review\/}~{\em 136}, 2608--2619.

\bibitem[\protect\citeauthoryear{Hamill}{Hamill}{2001}]{Hamill2001}
Hamill, T.~M. (2001).
\newblock {Interpretation of rank histograms for verifying ensemble forecasts}.
\newblock {\em Monthly Weather Review\/}~{\em 129}, 550--560.

\bibitem[\protect\citeauthoryear{Hamill and Colucci}{Hamill and
  Colucci}{1997}]{Hamill1997}
Hamill, T.~M. and S.~J. Colucci (1997).
\newblock {Verification of Eta-RSM short-range ensemble forecasts}.
\newblock {\em Monthly Weather Review\/}~{\em 125}, 1312--1327.

\bibitem[\protect\citeauthoryear{Hering and Genton}{Hering and
  Genton}{2010}]{HeringGenton2010}
Hering, A. and M.~G. Genton (2010).
\newblock {Powering up with space-time wind forecasting}.
\newblock {\em Journal of the American Statistical Association\/}~{\em 105},
  92--104.

\bibitem[\protect\citeauthoryear{Kann, Wittmann, Wang, and Ma}{Kann
  et~al.}{2009}]{Kann2009}
Kann, A., C.~Wittmann, Y.~Wang, and X.~Ma (2009).
\newblock {Calibrating 2-m temperature of limited area ensemble forecasts using
  high-resolution analysis}.
\newblock {\em Monthly Weather Review\/}~{\em 137}, 3373--3387.

\bibitem[\protect\citeauthoryear{Kleiber, Raftery, Baars, Gneiting, Mass, and
  Grimit}{Kleiber et~al.}{2011}]{Kleiber2011}
Kleiber, W., A.~E. Raftery, J.~Baars, T.~Gneiting, C.~F. Mass, and E.~Grimit
  (2011).
\newblock Locally calibrated probabilistic temperature forecasting using
  geostatistical model averaging and local {B}ayesian model averaging.
\newblock {\em Monthly Weather Review\/}~{\em 139}, 2630--2649.

\bibitem[\protect\citeauthoryear{Krzysztofowicz and Toth}{Krzysztofowicz and
  Toth}{2008}]{Krzysztofowicz2008}
Krzysztofowicz, R. and Z.~Toth (2008).
\newblock Bayesian processor of ensemble {(BPE)}: {C}oncept and implementation.
\newblock Workshop slides 4th {NCEP/NWS Ensemble User Workshop}, {Laurel},
  {Maryland}.
\newblock Available at
  \url{http://www.emc.ncep.noaa.gov/gmb/ens/ens2008/Krzysztofowicz_Pres}
  \url{entation_Web.pdf}.

\bibitem[\protect\citeauthoryear{Leutbecher and Palmer}{Leutbecher and
  Palmer}{2008}]{Leutbecher2008}
Leutbecher, M. and T.~N. Palmer (2008).
\newblock {Ensemble forecasting}.
\newblock {\em Journal of Computational Physics\/}~{\em 227}, 3515--3539.

\bibitem[\protect\citeauthoryear{Marquis, Wilczak, Ahlstrom, Sharp, Stern,
  Smith, and Calvert}{Marquis et~al.}{2011}]{Marquis2011}
Marquis, M., J.~Wilczak, M.~Ahlstrom, J.~Sharp, A.~Stern, C.~J. Smith, and
  S.~Calvert (2011).
\newblock {Forecasting the wind to reach significant penetration levels of wind
  energy}.
\newblock {\em Bulletin of the American Meteorological Society\/}~{\em 92},
  1159--1171.

\bibitem[\protect\citeauthoryear{{National Weather Service}}{{National Weather
  Service}}{1998}]{Service1998}
{National Weather Service} (1998).
\newblock {Automated Surface Observing System (ASOS) User's Guide}.
\newblock Available at \url{http://www.weather.gov/asos/aum-toc.pdf}.

\bibitem[\protect\citeauthoryear{Pinson}{Pinson}{2011}]{Pinson2011}
Pinson, P. (2011).
\newblock {Adaptive calibration of $(u ,v)$-wind ensemble forecasts}.
\newblock {\em Quarterly Journal of the Royal Meteorological Society, {\em in
  press}\/}.

\bibitem[\protect\citeauthoryear{Pinson and Hagedorn}{Pinson and
  Hagedorn}{2011}]{PinsonHagedorn2011}
Pinson, P. and R.~Hagedorn (2011).
\newblock Verification of the {ECMWF} ensemble forecasts of wind speed against
  analyses and observations.
\newblock {\em Meteorological Applications, {\em in press}\/}.

\bibitem[\protect\citeauthoryear{{R Development Core Team}}{{R Development Core
  Team}}{2011}]{R2011}
{R Development Core Team} (2011).
\newblock {\em R: A Language and Environment for Statistical Computing}.
\newblock Vienna, Austria: R Foundation for Statistical Computing.
\newblock {ISBN} 3-900051-07-0.

\bibitem[\protect\citeauthoryear{Raftery, Gneiting, Balabdaoui, and
  Polakowski}{Raftery et~al.}{2005}]{Raftery2005}
Raftery, A.~E., T.~Gneiting, F.~Balabdaoui, and M.~Polakowski (2005).
\newblock {Using Bayesian model averaging to calibrate forecast ensembles}.
\newblock {\em Monthly Weather Review\/}~{\em 133}, 1155--1174.

\bibitem[\protect\citeauthoryear{Roulston and Smith}{Roulston and
  Smith}{2003}]{Roulston2003}
Roulston, M. and L.~Smith (2003).
\newblock {Combining dynamical and statistical ensembles}.
\newblock {\em Tellus A\/}~{\em 55}, 16--30.

\bibitem[\protect\citeauthoryear{Schefzik}{Schefzik}{2011}]{Schefzik2011}
Schefzik, R. (2011).
\newblock {\em {Ensemble Copula Coupling}}.
\newblock Diploma thesis, Faculty of Mathematics and Informatics, University of
  Heidelberg.

\bibitem[\protect\citeauthoryear{Sch\"olzel and Friederichs}{Sch\"olzel and
  Friederichs}{2008}]{Schoelzel2008}
Sch\"olzel, C. and P.~Friederichs (2008).
\newblock Multivariate non-normally distributed random variables in climate
  research -- {I}ntroduction to copulas.
\newblock {\em Nonlinear Processes in Geosciences\/}~{\em 15}, 761--772.

\bibitem[\protect\citeauthoryear{Schuhen}{Schuhen}{2011}]{Schuhen2011}
Schuhen, N. (2011).
\newblock {\em {Ensemble Model Output Statistics for Wind Vectors}}.
\newblock Diploma thesis, Faculty of Mathematics and Informatics, University of
  Heidelberg.

\bibitem[\protect\citeauthoryear{Sloughter}{Sloughter}{2009}]{Sloughter2009}
Sloughter, J.~M. (2009).
\newblock {\em {Probabilistic Weather Forecasting Using Bayesian Model
  Averaging}}.
\newblock Ph.{D}.~thesis, Department of Statistics, University of Washington.

\bibitem[\protect\citeauthoryear{Sloughter, Gneiting, and Raftery}{Sloughter
  et~al.}{2010}]{Sloughter2010}
Sloughter, J.~M., T.~Gneiting, and A.~E. Raftery (2010).
\newblock {Probabilistic wind speed forecasting using ensembles and Bayesian
  model averaging}.
\newblock {\em Journal of the American Statistical Association\/}~{\em 105},
  25--35.

\bibitem[\protect\citeauthoryear{Sloughter, Gneiting, and Raftery}{Sloughter
  et~al.}{2011}]{Sloughter2011}
Sloughter, J.~M., T.~Gneiting, and A.~E. Raftery (2011).
\newblock {Probabilistic wind vector forecasting using ensembles and Bayesian
  model averaging}.
\newblock {\em Monthly Weather Review, {\em submitted}\/}.

\bibitem[\protect\citeauthoryear{Talagrand, Vautard, and Strauss}{Talagrand
  et~al.}{1997}]{Talagrand1997}
Talagrand, O., R.~Vautard, and B.~Strauss (1997).
\newblock {Evaluation of probabilistic prediction systems}.
\newblock In {\em Proc. Workshop on Predictability}, pp.\  1--25. Reading, UK,
  European Centre for Medium-Range Weather Forecasts.

\bibitem[\protect\citeauthoryear{Thorarinsdottir and Gneiting}{Thorarinsdottir
  and Gneiting}{2010}]{Thorarinsdottir2010}
Thorarinsdottir, T.~L. and T.~Gneiting (2010).
\newblock {Probabilistic forecasts of wind speed: Ensemble model output
  statistics by using heteroscedastic censored regression}.
\newblock {\em Journal of the Royal Statistical Society Series A\/}~{\em 173},
  371--388.

\bibitem[\protect\citeauthoryear{Thorarinsdottir and Johnson}{Thorarinsdottir
  and Johnson}{2011}]{Thorarinsdottir2011}
Thorarinsdottir, T.~L. and M.~S. Johnson (2011).
\newblock {Probabilistic wind gust forecasting using non-homogeneous Gaussian
  regression}.
\newblock {\em Monthly Weather Review, {\em in press}\/}.

\bibitem[\protect\citeauthoryear{Vardi and Zhang}{Vardi and
  Zhang}{2000}]{Vardi2000}
Vardi, Y. and C.-H. Zhang (2000).
\newblock {The multivariate $L_1$-median and associated data depth}.
\newblock {\em Proceedings of the National Academy of Sciences of the United
  States of America\/}~{\em 97}, 1423--1426.

\bibitem[\protect\citeauthoryear{Whitaker and Loughe}{Whitaker and
  Loughe}{1998}]{Whitaker1998}
Whitaker, J.~S. and A.~F. Loughe (1998).
\newblock {The relationship between ensemble spread and ensemble mean skill}.
\newblock {\em Monthly Weather Review\/}~{\em 126}, 3292--3302.

\bibitem[\protect\citeauthoryear{Wilks}{Wilks}{2009}]{Wilks2009}
Wilks, D.~S. (2009).
\newblock {Extending logistic regression to provide
  full-probability-distribution MOS forecasts}.
\newblock {\em Meteorological Applications\/}~{\em 16}, 361--368.

\bibitem[\protect\citeauthoryear{Wilks}{Wilks}{2011}]{Wilks2011}
Wilks, D.~S. (2011).
\newblock {\em {Statistical Methods in the Atmospheric Sciences}\/} (third
  ed.).
\newblock Academic Press.

\bibitem[\protect\citeauthoryear{Wilks and Hamill}{Wilks and
  Hamill}{2007}]{WilksHamill2007}
Wilks, D.~S. and T.~M. Hamill (2007).
\newblock {Comparison of ensemble-MOS methods using GFS reforecasts}.
\newblock {\em Monthly Weather Review\/}~{\em 135}, 2379--2390.

\end{thebibliography}
\bibliographystyle{chicago}

\end{document}